\documentclass[adraft,copyright,creativecommons]{eptcs}
\providecommand{\event}{SOS 2007} 
\usepackage{breakurl}             
\usepackage{underscore}           

\usepackage{amsfonts}

\usepackage{amsfonts,amstext,amsmath}

\usepackage{xparse}  
\usepackage{xifthen} 
\usepackage{xspace}
\usepackage{stackengine}
\usepackage{mathtools}

\usepackage{amsthm}

\newtheorem{theorem}{Theorem}

\theoremstyle{definition}
\newtheorem{defn}{Definition}[section]

\newcommand{\defeq}{\stackrel{\operatorname{\scriptscriptstyle def}}{\raisebox{-1.0pt}{$=$}}}

\renewcommand{\S}{\mathcal{S}}

\DeclareDocumentCommand{\pred}{O{}O{}mO{}}{%
  \ifthenelse{\isempty{#4}}{\textrm{Pre}_{#2}^{#1}\left(#3\right)}{\textrm{Pre}_{#2}^{#1}\left(#3, #4\right)}
}

\DeclareMathOperator{\upc}{
  \mathchoice{\uparrow}
             {\uparrow}
             {\uparrow \!}
             {\uparrow \!}
}

\DeclareDocumentCommand{\trans}{O{}O{}m}{%
  \xrightarrow{#3}\mathrel{\vphantom{\to}^{#1}_{\ifthenelse{\isempty{#2}}{}{#2}}}
}

\DeclareDocumentCommand{\ctrans}{O{}O{}m}{%
  \overset{\mathclap{#3}}{\leadsto}\mathrel{\vphantom{\to}^{#1}_{\ifthenelse{\isempty{#2}}{}{#2}}}
}

\DeclareDocumentCommand{\auttrans}{O{}m}{%
  \overset{\scriptstyle\mathclap{#2}}{\dashrightarrow}\mathrel{\vphantom{\to}_{\ifthenelse{\isempty{#1}}{}{#1}}}
}

\DeclareDocumentCommand{\succ}{O{}O{}mO{}}{%
  \ifthenelse{\isempty{#4}}{\textrm{Post}_{#2}^{#1}\left(#3\right)}{\textrm{Post}_{#2}^{#1}\left(#3, #4\right)}
}

\DeclareDocumentCommand{\ssucc}{O{}mO{}}{%
  \succ[#1][\S]{#2}[#3]
}

\title{From Well Structured Transition Systems to Program Verification}
\author{Alain Finkel
\institute{Universit\'e Paris-Saclay, ENS Paris-Saclay, CNRS,}
\institute{Laboratoire
Sp\'ecification et V\'erification, 91190, Gif-sur-Yvette, France.}
\institute{Institut Universitaire de France.}
\email{finkel@lsv.fr}
}
\def\titlerunning{From Well Structured Transition Systems to Program Verification}
\def\authorrunning{Alain Finkel}
\begin{document}
\maketitle
\noindent
{\bf Abstract:}
We describe the use of the theory of WSTS for verifying programs.\\
%

\section{Preliminaries}

A relation $\leq\ \subseteq X \times X$ over a set $X$ is
a \emph{quasi-ordering} if it is reflexive and transitive, and a
\emph{partial ordering} if it is antisymmetric as well.  It is
\emph{well-founded} if it has no infinite descending chain.  A
quasi-ordering $\leq$ is a \emph{well-quasi-ordering} (resp.\
\emph{well partial order}), \emph{wqo} (resp.\ \emph{wpo}) for short,
if for every infinite sequence $x_0, x_1, \dots \in X$, there exist
$i < j$ such that $x_i \leq x_j$.  This is strictly stronger than
being well-founded.

One example of well-quasi-ordering is the componentwise ordering of
tuples over $\mathbb{N}$. More formally, $\mathbb{N}^d$ is well-quasi-ordered by
$\leq$ where, for every $\textbf{x}, \textbf{y} \in \mathbb{N}^d$,
$\textbf{x} \leq \textbf{y}$ if and only if $\textbf{x}(i) \leq \textbf{y}(i)$ for
every $i \in [d]$. We extend $\mathbb{N}$ to
$\mathbb{N}_{\omega} \defeq \mathbb{N} \cup \{\omega\}$ where $n \leq \omega$ for every
$n \in \mathbb{N}_{\omega}$. $\mathbb{N}_{\omega}^d$ ordered componentwise is also
well-quasi-ordered. Let $\Sigma$ be a finite alphabet. We write
$\Sigma^*$ to denote the set of finite
words over $\Sigma$. 
For every $u,
v \in \Sigma^*$, we write $u \sqsubseteq v$ if $u$ is a subword of
$v$, i.e. $u$ can be obtained from $v$ by removing zero, one or
multiple letters. $\Sigma^*$ is well-quasi-ordered by $\sqsubseteq$.

\section{Well Structured Transition Systems}

\subsection{Well structured transition systems: wqo and monotony}

An \emph{ordered (labeled) transition system} is a triple $(X,
\trans{\Sigma}, \leq)$ such that $(X, \trans{\Sigma})$ is a
\emph{(labeled) transition system} and $\leq$ is a quasi-ordering. An
ordered transition system $\S$ is a \emph{well structured transition
  system (WSTS)} if $\leq$ is a well-quasi-ordering and $\S$ is
\emph{monotone}, i.e. for all $x, x', y \in X$ and $a \in \Sigma$ such
that $x \trans{a} y$ and $x' \geq x$, there exists $y' \in X$ such
that $x' \trans{*} y'$ and $y' \geq y$. Many other types of
monotonicities were defined in the literature (see ~\cite{finkel98b}),
but, for our purposes, we only need to introduce strong
monotonicities. We say that $\S$ has \emph{strong monotonicity} if for
all $x, x', y \in X$ and $a \in \Sigma$, $x \trans{a} y$ and $x' \geq
x$ implies $x' \trans{a} y'$ for some $y' \geq y$. We say that $\S$
has \emph{strong-strict monotonicity}\footnote{Strong-strict
  monotonicity should not be confused with strong \emph{and} strict
  monotonicities. Here strongness and strictness have to hold at the
  \emph{same} time.} if it has strong monotonicity and for all $x, x',
y \in X$ and $a \in \Sigma$, $x \trans{a} y$ and $x' > x$ implies $x'
\trans{a} y'$ for some $y' > y$.

\begin{theorem}\label{wsts}\cite{F-icalp87,finkel98b,DBLP:journals/iandc/AbdullaCJT00}
Termination, boundedness, control-state reachability and coverability are decidable for effective WSTS with strong-strict monotony.
\end{theorem}

There are two main techniques for proving these decidability results: backward and forward analysis. The backward coverability algorithm allows to compute the finite basis of the set of all predecessors of the upward closure of a state. The forward coverability algorithm computes the finite reduced reachability tree and the finite (extended) Karp-Miller tree (under supplementary hypothesis): these two forward algorithms operate with inductive downward closed invariants.

\subsection{A short story of well structured transition systems}

\noindent

\emph{Well structured transition systems} (initially called \emph{structured transition systems}  in \cite{F-icalp87}) were initially defined and
studied as monotone transition systems
equipped with a well-quasi-ordering on their set of
states. Termination was shown decidable for \emph{well structured
  transition systems} with \emph{transitive} monotonicity, while
boundedness was shown decidable for well structured transition systems with
\emph{strict} monotonicity  in \cite{F-icalp87}. For a subclass of finitely branching
labeled well structured transition systems with strong-strict monotonicity,
now called \emph{very well structured transition systems} in
\cite{BFG-fsttcs17}, a generalization of the Karp-Miller algorithm
was shown to compute their coverability sets \cite{F-icalp87,BFG-fsttcs17}. In \cite{DBLP:journals/iandc/AbdullaCJT00},
the coverability problem was shown to be decidable for a subclass of
well structured transition systems,
i.e. \emph{labeled} well structured transition systems with \emph{strong
  monotonicity} ~\cite[Def.~3.4]{DBLP:journals/iandc/AbdullaCJT00} and satisfying an additional \emph{effective
  hypothesis}: the existence of an algorithm to compute the finite set $min(\pred{\upc{s}})$
of minimal elements of $\pred{\upc{s}}$, where $\pred{\upc{s}}$ is
the set of immediate predecessors of the upward-closure $\upc{s}$ of a
state $s$.
 In \cite{finkel98b}, mathematical properties were
distinguished from effective properties, and the coverability problem
was shown decidable for the \emph{entire} class of well structured
transition systems satisfying the similar additional \emph{effective
  hypothesis} that there exists an algorithm to compute the finite set
$min(\upc{\pred{\upc{s}}})$, i.e., the hypotheses of transitions
labeling and strong monotonicity made in \cite{DBLP:journals/iandc/AbdullaCJT00} turned out to
be superfluous.

Today, following the presentation of~\cite{finkel98b}, what is
\emph{mathematically} known as \emph{well structured transition
  systems} (or shortly \emph{well structured
  systems}) is exactly the original class of \emph{structured
  transition systems} \cite{F-icalp87}; and necessary effective hypotheses are added
for obtaining decidability of properties such as termination,
control-state reachability, coverability and boundedness.

\section{From Programs to Well Structured Transition Systems} 
\subsection{The general method}
Given a program $P$ and a safety property $\phi$, let's describe two steps for verifying that $P$ satisfies $\phi$ by using WSTS:
\begin{enumerate}
\item The first step is to build a transition system $(S,\rightarrow)$ associated with $(P,\phi)$. This is well known as the operationnal semantics of the program and we are used to this. But the problem is the hudge size of the associated transition system. In general we will define and compute an abstraction of the original program $P$ because we may (and must) forget some useless parts of the program that have no effect on property $\phi$. A kind of such activities is the (static and dynamic) slicing that computes parts of the program that may modify a set of variables and this computation can be done with a small cost. There exist other techniques to build abstractions of the program that produce smaller and tractable programs.  We have also to translate the property $\phi$ on $P$ into a state-property $\phi_S$ in $(S,\rightarrow)$ (sometimes a formula in a logic) that would be decidable for WSTS. 

\item The second step is to look for an ordering $\leq$ having these two desired properties (monotony and well ordering), i.e., such that $(S,\rightarrow, \leq)$ is WSTS. Let us recall that the termination ordering makes of each transition system a WSTS \cite{finkel98b} but this ordering is undecidable so the obtained WSTS is not effective and we cannot deduce the decidability of usual properties. If we find such decidable ordering $\leq$, we just verify whether $(S,\rightarrow, \leq)$ satisfies the state-property $\phi_S$. To make this verification, one usually reduces $\phi_S$ to a coverability property in $(S,\rightarrow, \leq)$.
\end{enumerate}

\subsection{What can you do when you can't find a monotone well ordering ?}

Let us analyse two cases that are not directly translatable into WSTS. 

\subsubsection{We found a well ordering which is not strongly monotone}

Let us consider the case in which we found a well ordering $\leq$ but $(S,\rightarrow,\leq)$ is unfortunately not strongly monotone. Apart from the usual well ordering on integers (Dickson), there exist many well orderings on different kinds of sets: let us enumerate, the multiset ordering, the subword ordering on finite words (Higman), the homeomorphic embedding on finite trees (Kruskal), the minor ordering (Robertson $\&$ Seymour) on finite graphs,...etc. These orderings can be often extended to the infinite. With Jean Goubault-Larrecq, we define in \cite{DBLP:conf/stacs/FinkelG09} an algebra allowing the composition of well orderings by many operators like finite cartesian product. 

Let us consider a counter machine $M$. Recall that the usual ordering on positive integers (which extends to vectors of integers) is well (Dickson Lemma) but it is not (strongly) monotone on general counters machines because the guards containing tests to zero are typically not monotone. 
We may change the original machine into another one which will be a WSTS. We may change the operations and/or the states.

A first drastic action is to remove the tests to zero; another possibility is to replace tests to zero by resets (or by transfers). The new machine $M_{new}$ is now monotone, hence machine $M_{new}$ is a WSTS (for the usual ordering) that over-approximates the original counter machine $M$. If $M_{new}$ never meets a bad state then one may deduce the same for $M$. Other properties like termination, boundedness, non-reachability are also preserved by monotonic abstraction \cite{DBLP:journals/ijfcs/AbdullaDHR09}.

We may change the states by abstracting them modulo an equivalence relation $\equiv$ or even with an ordering.
One may also look for a computable abstraction $(S',\rightarrow',\leq')$ of $(S,\rightarrow,\leq)$ where $S'=(S/\equiv)$ and $\leq'=(\leq/\equiv)$ are an abstraction of $(S,\leq)$ such that the new transition relation $\rightarrow'$ (between abstract states in $S'$) is monotone with respect to $\leq'$ which must be still well and then $(S',\rightarrow',\leq')$ is a WSTS. The Abstract Interpretation \cite{DBLP:conf/popl/CousotC77} could be completed in the direction to produce WSTSs.

Another way is to consider general non monotone models and to test if a particular instance of the model is strongly monotone. This question is decidable, for example, for Presburger counter machine \cite{GF-fsttcs19}.

\subsubsection{We found a strongly monotone ordering which is not well}

A first possibility is use algorithms in WSTS as semi-algorithms in strongly monotone transition systems. But there is another way. The ordering which is not well on the considered set of states could be well on the subset of reachable states. In general, the reachability set is not computable but in some cases, it is possible to compute an overapproximation of the reachability set on which the ordering is well.

Another way is to consider general strongly monotone non-well ordered transition systems and to test if a particular ordering is well. This question is decidable, for example, for orderings defined by Presburger formulas (Presburger orderings) (see \cite{GF-fsttcs19} for the decidability for orderings in $\mathbb{N}$).

\section{Examples} 

\subsection{Programs with integers} 

Many programs can be modeled as counter machines (for example programs with lists \cite{BFLS-avis06}). Presburger counter machines (PCM) are a general model that allows to express guards and operations as Presburger formulas. It is clear that PCM contain Minsky machines and, as an immediate consequence, all non-trivial properties are undecidable for PCM. Let us now illustrate some notions introduced in step $2$ of the strategy described before. Let $M=(Q,...)$ be a Presburger counter machine with a set finite set $Q$ of control-states and $d$ counters. Let us first consider the most natural well ordering $\leq$ on integers that we classically extend on vectors as follows: let $\preceq  \ \defeq \ =_Q \times \leq^d$ where $=_Q$ is the equality on the finite set $Q$ and $ \leq^d$ is the vector ordering component by component. By Dickson Lemma, we know that $\preceq$ is still well. We cannot directly decide whether $M$ is strongly monotone for $\preceq$ but we may decide the strong monotony property for $M$ because both the description of $M$ and of the strong monotony property can be expressed as Presburger formulas  \cite{GF-fsttcs19}. If $M$ is strongly monotone for $\preceq$, we may use the WSTS theory. In the case where $M$ is not strongly monotone for $\preceq$, we may use the following (non-terminating) semi-algorithm that enumerates Presburger formulas $\psi_1, \psi_2,....,\psi_n,...$ representing well orderings $\leq_1, \leq_2,...,\leq_n,...$ on $\mathbb{N}^d$ and test, for all $n$, whether $M$ is strongly $\leq_n$-monotone. If there exists an integer $n \geq 1$ such that $\leq_n$ is well and strongly monotone on $M$, then the termination of the previous semi-algorithm is insured. But if there don't exist such $n$, this enumeration will never terminate and then it don't provide an algorithm to decide whether there exists a strongly monotone Presburger well ordering for $M$.
Let us define the class of \emph{existentially (strongly) well structured} Presburger counter machines as follows:

\begin{defn}
A Presburger counter machine $M$ is \emph{existentially well structured (resp. existentially strongly well structured)} if there exists a Presburger well ordering that is monotone (resp. strongly monotone) for $M$. 
\end{defn}

Coverability and other properties (see Theorem \ref{wsts}) are decidable for existentially well structured PCMs.
We may prove that the monotony property is undecidable \cite{GF-fsttcs19} for PCM of dimension one (and for Minsky machines of dimension $2$) with the usual well ordering on integers and we conjecture that the \emph{existentially well structured problem} (i.e., whether a PCM is existentially well structured) is also undecidable. Another natural (and still open) question is then to know whether the \emph{existential strongly well structured problem} is decidable for PCMs.
%
%

%

\subsection{Communication protocols} 

Let us consider a distributed program composed of a finite set of processes (finite automata, pushdown processes,...) that exchanges messages through fifo channels. We know that queue automata also called fifo machines (i.e., a finite automaton that communicates with an unique fifo buffer also called a \emph{bi-directional} fifo channel) may simulate Turing machines and counter machines \cite{DBLP:journals/tcs/VauquelinF80} and this is still true for two finite automata communicating through \emph{one-directional} fifo channels \cite{DBLP:journals/jacm/BrandZ83}. Let us consider, for simplifying notations, fifo machines (a single sequential control-graph) $M=(Q,...)$ communicating with $d$ channels and the most natural ordering on words, adapted to the fifo behavior, say the prefix ordering $\leq_{prefix}$ that is extended as previously by $\preceq_{prefix}  \ \defeq \ =_Q \times \leq_{prefix}^d$. Unfortunatly this ordering is not monotone neither well (except in the trivial case where the channel alphabets are reduced to an unique letter). The subword ordering $\sqsubseteq$ on finite words is well (Higman's Theorem) and its classical extension $\preceq_{\sqsubseteq}  \ \defeq \ =_Q \times \sqsubseteq^d$ is also well but it is not monotone on fifo machines ; however, $\preceq_{\sqsubseteq}$ is monotone on fifo machines with other semantics (like lossy, insertion), hence such non-perfect fifo machines are WSTS for the extended subword ordering. These kind of non-perfect fifo machines over-approximates original perfect fifo machines and we may apply the monotonic abstraction described previously in Section $3$. 

\subsection{Other programs} 

There exist many other illustrations of the power of WSTS to verify programs like hardware design, multithreaded programs, distributed systems. Let's quote programs with pointers and the use of graphs and orderings on graphs (subgraph ordering and minor ordering) to model the state of the memory \cite{DBLP:conf/atva/AbdullaACJ09}, parameterized verification of distributed algorithms \cite{DBLP:conf/rp/DelzannoS14}, programs with time constraints (timed Petri nets), cryptographic protocols \cite{DBLP:journals/corr/abs-1911-05430}, broadcast protocols,...etc.
%

\nocite{*}
\bibliographystyle{eptcs}
\bibliography{biblio-vpt}
\end{document}